\documentclass[reprint,amsmath,amssymb,prl,floatfix]{revtex4-1}

\usepackage{graphicx}
\usepackage{dcolumn}
\usepackage{bm}
\usepackage{amsmath}
\usepackage{siunitx}
\usepackage{epstopdf}

\begin{document}

\preprint{APS/123-QED}

\title{Magneto-transport controlled by Landau polariton states}

\author{Gian L. Paravicini-Bagliani$^{1,*}$, Felice Appugliese$^1$, Eli Richter$^1$, Federico Valmorra$^1$, Janine Keller$^1$, Mattias Beck$^1$, Nicola Bartolo$^3$, Clemens R\"ossler$^2$, Thomas Ihn$^2$, Klaus Ensslin$^2$,Cristiano Ciuti$^3$, Giacomo Scalari$^{1,\dagger}$ \& Jerome Faist$^1$}
\affiliation{$^1$Institute for Quantum Electronics, ETH Zurich, Auguste-Piccard-Hof 1, 8093 Zurich, Switzerland}%
\affiliation{$^2$Laboratory for Solid State Physics, ETH Zurich, Otto-Stern-Weg 1, 8093 Zurich, Switzerland}
\affiliation{$^3$Laboratoire Mat\'eriaux et Ph\'enom\`enes Quantiques, Universit\'e Paris Diderot, 10 Rue Alice Domon et L\'eonie Duquet, 75013 Paris-13E-Arrondissement, France}

\date{\today}

\maketitle

\textbf{Hybrid excitations, called polaritons, emerge in systems with strong light-matter coupling. Usually, they dominate the linear and nonlinear optical properties with applications
in quantum optics. Here, we show the crucial role of the electronic component of polaritons in the magneto-transport of a cavity-embedded 2D electron gas in the ultrastrong coupling regime. We show that the linear dc resistivity is significantly modified by the coupling to the cavity even without external irradiation. Our observations confirm recent predictions of vacuum-induced modification of the resistivity. Furthermore, photo-assisted transport in presence of a weak irradiation field at sub-THz frequencies highlights the different roles of localized and delocalized states.}

The strong light-matter coupling regime \cite{thompson1992observation,raimond2001manipulating} is realized  when the coupling $\Omega$ between photons and a material's excitation of frequency $\omega$ exceeds the losses $\gamma_{tot}$ of both components. An especially interesting situation is attained when quantum fluctuations of the electromagnetic field ground state give rise to the so-called vacuum Rabi splitting of the cavity polaritons. Solid-state systems \cite{weisbuch1992observation,dini2003microcavity,smolka2014cavity} have recently proven to be instrumental in achieving the ultimate limit of this kind of coupling.  The \textit{ultrastrong coupling} regime \cite{ciuti2005quantum,anappara2009signatures, gunter2009sub, niemczyk2010circuit, forn2010observation,todorov2010ultrastrong,jouy2011transition, muravev2011observation, geiser2012ultrastrong, scalari2012ultrastrong, muravev2013ultrastrong,gambino2014exploring,orgiu2015conductivity,zhang2016collective,paravicini2017gate}, realized in the limit of $\Omega/\omega \gtrsim 0.1$, exploits the collective nature of the matter excitations \cite{ciuti2005quantum,todorov2012intersubband,hagenmuller2010ultrastrong} to achieve a peculiar situation where the ground state of the system is constituted by non-trivial quantum vacua \cite{ciuti2005quantum}.

The (ultra-)strong coupling regime has so far mostly been investigated by interrogating the photonic component of the polariton quasi-particle weakly probing the coupled system with low photon fluxes \cite{thompson1992observation,raimond2001manipulating,weisbuch1992observation,dini2003microcavity,anappara2009signatures, niemczyk2010circuit, forn2010observation, muravev2011observation, geiser2012ultrastrong, scalari2012ultrastrong, muravev2013ultrastrong,smolka2014cavity,gambino2014exploring,zhang2016collective,sidler2017fermi,ravets2017polaron,samkharadze2018strong}. Notable exceptions have been the measurements of the matter part of an exciton polariton condensate with an excitonic 1s-2p transition\cite{Menard2014Revealing} and a transport experiment in molecules coupled to a plasmonic resonance \cite{orgiu2015conductivity}.

Recently we pioneered a new experimental platform, the Landau polaritons, to study ultrastrong light matter interactions \cite{scalari2012ultrastrong,maissen2014ultrastrong}  allowing to reach record-high normalized light-matter coupling ratios $\Omega/\omega_{cav} > 1$ \cite{bayer2017terahertz}. The inter-Landau level (cyclotron) transition $\omega_c=\frac{eB}{m^*}$ ($m^*$: effective electron mass) of a two-dimensional electron gas (2DEG) under strong magnetic field  is coupled to a complementary electronic LC resonator \cite{chen:OE:07} at frequencies of 100's of GHz, which plays effectively the role of the optical cavity. This system is especially well suited to study the matter part of ultrastrongly coupled polaritons using low temperature magneto-transport.

\begin{figure*}[!ht]
 \centering
 \includegraphics[width=\textwidth]{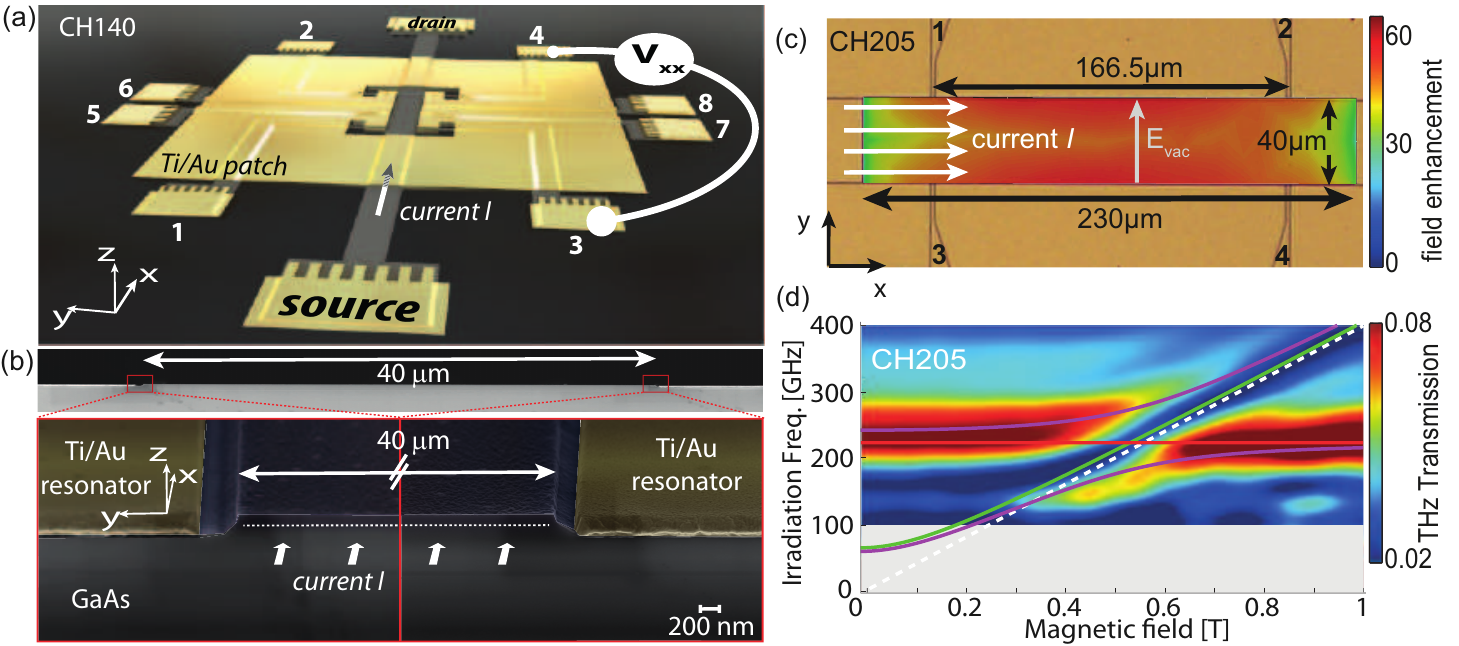}
 \caption{\label{figure1} \textbf{Quantum Hall system ultrastrongly coupled to a microvave cavity}
a) Sample schematic: An AC current ($I=100$ nA) is applied in x-direction along a 40 $\mu m$ wide GaAs/AlGaAs Hall bar between source (S) and drain (D) contacts. `CH140' sketched here shows voltage probes located entirely inside the region where the vacuum field created by the LC-like resonator (patterned patch of gold around Hall bar) has its maximum.
b) (top) SEM picture of a y-z cross-section across the Hall bar channel. (bottom) Zoom-in on the edges of the Hallbar showing the capacitive gap of the resonator (yellow) formed by a Ti/Au-layer very closely surrounding both sides of the Hall bar without covering it.
c)  Detail micrograph of second resonator Hallbar `CH205' with the simulated $\lambda/2$-like mode distribution at 205 GHz overlayed as colormap (see supplementary for details).
d) Free space THz transmission trough a sample featuring an array of Hallbars of type CH205  ($T_{sample}=2.9$ K). Magneto plasmon polariton dispersion fits \cite{paravicini2017gate} are overlayed as magenta curves. They result from the anti-crossing of the magneto-plasmon dispersion (green) and the resonator frequency (red). The computed cyclotron dispersion is shown in white. We observe a normalized light-matter coupling ratio $\Omega/\omega_{cav}=30\%$ and 20 $\%$ for CH140 and CH205 respectively. The corresponding figures for CH140 are shown in the supplementary Figs.~S1b) and c)}
\end{figure*}

It was recently proposed theoretically \cite{bartolo2018vacuum} that such transport is actually driven by the {\em bright} polariton operator, i.e. the same operator driving the optical response.  Here we find experimental evidence consistent with this picture, in which most tellingly the longitudinal resistivity $\rho_{xx}$ bears the signatures of the polariton branches.

Further confirmation for polaritonic effects acting on magneto-transport is obtained by observing the resistance change $\Delta\rho_{xx}$ induced by an extremely weak excitation of the system with a tunable narrow band source. Exploiting the extremely large and filling factor dependent responsivity of 2DEG in the THz\cite{maan1982observation,kawano2001highly,dorozhkin2007coexistence}, we reveal the response of the polariton state in a small energy slice $\Delta E \approx k_B T_{el}$ around the Fermi energy.

A sample schematic in Fig.~\ref{figure1}a) and an SEM picture of the sample cross section in Fig.~\ref{figure1}b) show the Hallbar very closely surrounded by the capacitive gap of the metallic resonator. We use two different geometries displaying resonant field enhancement at 140 GHz and 205 GHz (referred to as `CH140' and `CH205', respectively, shown in Figs.~\ref{figure1}a) and c)).  As a reference, we use a completely uncovered Hall bar `RH' without a resonator.
A finite element (FE) simulation (CST microwave studio) of the electromagnetic mode distribution of CH205 is overlayed as colormap in Figs.~\ref{figure1}c). The colormap shows the resonator's in-plane electric field distribution $E_{x,y}=\sqrt{|E_x|^2+|E_y|^2}$ scaled as the local field enhancement. Both cavities create a strong vacuum electric field $\vec{E}_{vac}\perp \vec{I}$ in the entire area spanned by the Hall bar contacts 1 to 4.

The THz transmission measurement shown in Fig.~\ref{figure1}d) hence shows the anti-crossing behaviour of the electronic resonance (green curve) strongly coupled to the resonator (red). As discussed in the supplementary material, to correctly model the dispersion of the lower polariton branch at low magnetic field, the cyclotron dispersion was substituted by the one of magneto-plasmon\cite{stern1967polarizability, allen1977observation} originating from the lateral confinement of the electrons in the Hall bar. We find a normalized light-matter coupling  $\Omega/\omega_{cav}$ equal to 30$\%$ and 20$\%$ ($\pm 5 \%$) for CH140 and CH205, respectively. From the linewidth in Fig.~\ref{figure1}d), the cavities quality factor is $Q\sim5$. Finite element simulations shown in supplementary Figs.~S2a) to c) qualitatively reproduce the THz transmission results for the three samples.

\begin{figure*}[!ht]
 \centering
 \includegraphics[width=\textwidth]{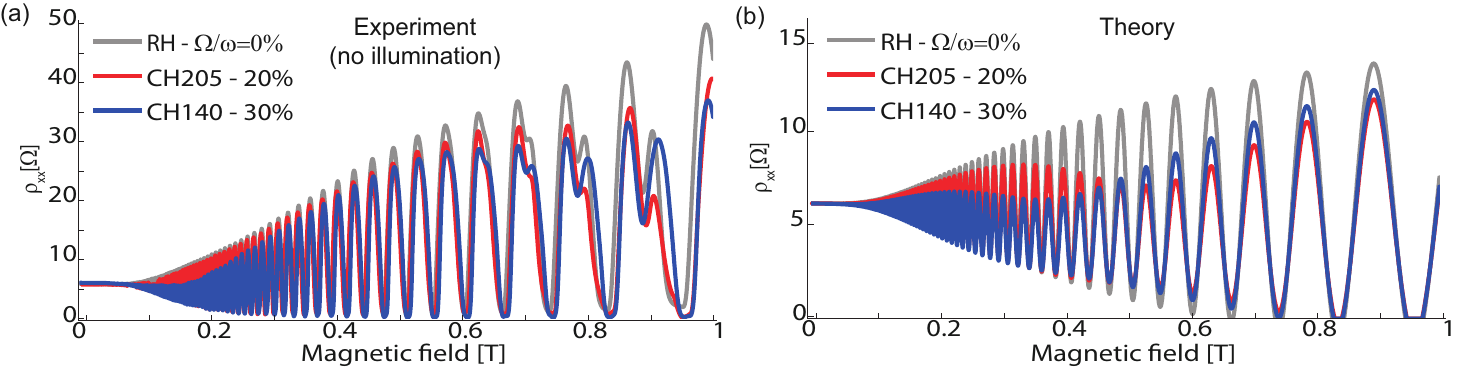}
 \caption{\label{figure2} \textbf{Transport in vacuum fields}
a) Longitudinal resistance $\rho_{xx}$ for the three Hallbars shows a decreasing Shubnikov-de Haas modulation amplitude with increasing light-matter coupling. ($T_{sample}=100$ mK, $\langle n_{polariton} \rangle<10^{-3}$)
b) Theoretically computed resistance\cite{bartolo2018vacuum} for 0$\%$(gray),20$\%$(red) and 30$\%$(blue) normalized light matter coupling corresponding to the 3 measurements in a). The traces qualitatively reproduce most features of the measurement in a).}
\end{figure*}

Fig.~\ref{figure2}a) shows the longitudinal resistance $\rho_{xx}=\frac{V_{xx}W}{I L}$ versus magnetic field for the three samples measured between contacts 1 and 2 (as marked in Fig.~\ref{figure1}c)). We observe the well known Shubnikov-de Haas oscillations. No THz illumination was used and the computed thermal photon population was negligable at the resonator frequency. Nevertheless, a clear dependence of the Shubnikov-de Haas amplitude is observed as a function of $\Omega/\omega$. The density $n_s=3.3\times 10^{11} cm^{-2}$ and mobility $\mu=3.1 \times 10^6 cm^2/Vs$ are identical within 2\% for all Hall bars despite the presence of the metal around the resonator Hall bars. 
For comparison, Fig.~\ref{figure2}b) shows computed resistance traces in the presence of the polaritonic vacuum\cite{bartolo2018vacuum}. Apart from the cavity related scattering $\tau_p=300 ps$, all parameters are fixed by the experiment. The theoretical curves in Fig.~\ref{figure2}b) strikingly reproduce most qualitative features found in the measurement in Fig.~\ref{figure2}a), showing the strong dependence of the magneto-transport on the cavity polaritons.

In a second set of experiments, we study the influence of the polariton population on the transport, by weakly illuminating the sample with tunable, narrow-band sub-THz radiation. The longitudinal resistance under weak illumination $\rho_{xx}^{illu}(B,\omega_{irr})$ is obtained by tuning the single frequency source from 60 GHz up to 600 GHz, while keeping the magnetic field fixed. Such a frequency sweep is repeated for different values of the magnetic field scanning it in small steps. From the irradiation power we estimate that $\sim 6$ polariton excitations are present in the system, which is a small fraction $\sim 2\times 10^{-5}$ of the available electrons in the highest Landau level (see supplementary material).

\begin{figure}[!ht]
 \centering
 \includegraphics[width=\columnwidth]{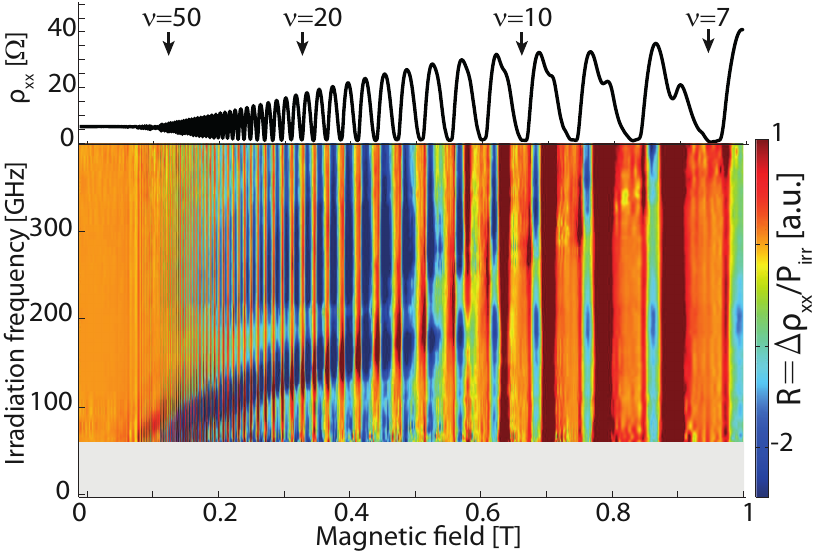}
 \caption{\label{figure3} \textbf{Transport with small polariton population}
Colormap: Irradiation-induced change of the longitudinal resistance $\rho_{xx}$ of CH205 normalized to irradiation power $R(B,\omega_{irr})$ as a function of irradiation frequency and magnetic field. The black trace above shows the dark (no microwave irradiation) trace of $\rho_{xx}$ (reprint from Fig.~\ref{figure2}a). In contrast to THz-transmission in Fig.~\ref{figure1}d), the resistance change depends strongly on the value of the resistance $\rho_{xx}$, thus on the filling factor $\nu$ marked with black arrows ($T_{sample}=100$ mK).}
\end{figure}

As the illumination power $P_{irr}$ changes with illumination frequency $\omega_{irr}$ (see Fig.~S3), the colormap in Fig.~\ref{figure3} shows the photo-response under irradiation $R(B,\omega_{irr})=(\rho_{xx}^{illu}-\rho_{xx}^{dark})/P_{irr}$ of CH205 as a function of magnetic field and irradiation frequency. Like the longitudinal resistance $\rho_{xx}$, $R(B,\omega_{irr})$ oscillates in phase with the density of states (DOS) at the Fermi energy $E_F$ periodically with the (spin degenerate) filling factor $\nu=hn_s/(2eB)$. In contrast, the THz-transmission measured with THz time-domain spectroscopy through an array of the same 205 GHz cavities (Fig.~\ref{figure1}d)) is independent of the location of the Fermi level relative to the ladder of Landau energy levels $E_n$.

\begin{figure*}[!ht]
 \centering
 \includegraphics[width=\textwidth]{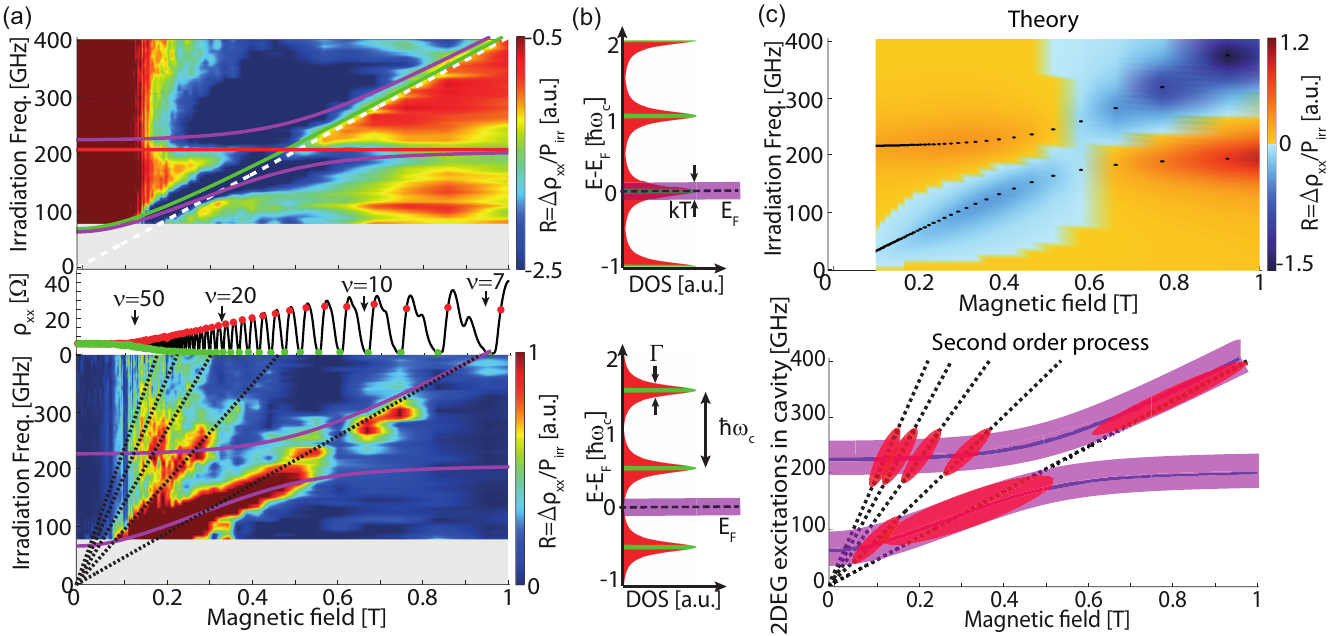}
 \caption{\label{figure4} \textbf{Filling-factor-dependent photo-response reveals polariton branches and its decay channels} 
a) CH205: The center panel shows the longitudinal resistance in the dark $\rho_{xx}^{dark}$ (black trace) with the filling factor marked with arrows. For the top and bottom panels, we select only the measurements marked in the center panel with red and green dots overlapped on the black trace. These approximately correspond to the half integer and integer filling factors respectively, where $\rho_{xx}^{dark}$ reaches its maxima and minima. (top) The longitudinal photo-response $R=\Delta\rho_{xx}/P_{irr}$ shows a change when resonant to the magneto plasmon polariton dispersions fitted with magenta curves (see supplementary material for the model\cite{paravicini2017gate}). Polaritons result from the anti-crossing of the resonator frequency (red) and the magneto-plasmon dispersion (green), which itself converges to the cyclotron dispersion (white dashed). When resonant to the polaritons, we have $\SI{-3}{\ohm}\lesssim \Delta \rho_{xx}\lesssim\SI{-1}{\ohm}$. (bottom) At integer filling factors with localized states at the Fermi energy, a set of linear dispersions appears (black dashed). These are attributed to the inter-Landau level transition ($\Delta \rho_{xx}\sim \SI{+1}{\ohm}$) and its higher orders ($\Delta \rho_{xx}\sim \SI{+0.3}{\ohm}$). See Figs.~S4b) and c) for RH and CH140.
b) Sketch of electronic density of states versus energy near the Fermi energy $E_F$. Landau levels are represented as Lorentzians with width $\Gamma(=\hbar/\tau_q)$ for the case when $E_F$ is inside one Landau Level (top panel) and exactly in between (bottom). Electrons are delocalized (localized) near the center of each level (green area). The electrons relevant to transport are distributed within a narrow range of width $k_BT$ around $E_F$ (magenta region). 
c) (top) Theoretically computed photo-response at half integer filling factors shows a response when resonant to the polariton branches as seen in the experiment.
(bottom) Sketch of non-radiative polariton decay process: Upon THz irradiation, polariton excitations are created, as shown by broad magenta lines.  These will decay non-radiatively most efficiently when they are energetically resonant with empty Landau levels above the Fermi energy (black lines). The intersection between these two sets of curves (red ovals) indicates the regions where the magneto resistance increases in the situation where the Fermi energy lies between two Landau levels (see Fig.~\ref{figure2}, lower panels).}
\end{figure*}

To highlight the filling factor dependence of the photo-response, we construct two color plots in Fig.~\ref{figure4}a) with two subsets of the `CH205'-data from Fig.~\ref{figure2}c), selecting only measurements at integer and half-integer filling factors, respectively, with spline interpolation in between.  (The complete data is shown in the supplementary material). The top panel in Fig.~\ref{figure4}a) shows measurements taken with $E_F=E_n$ ($\nu$ is half-integer). 
The bottom panel show measurements where $E_F=E_n+\frac{1}{2}\hbar\omega_c$ ($\nu$ is integer). The dark trace $\rho_{xx}^{dark}$ is shown in the central panel, with the red and green dots overlayed showing which measurements have been selected for the upper and lower panel respectively.

The response at half-integer filling factors (top panel) shows clear signatures of the cavity-coupled quasi-particle: a photo-response occurs when $\omega_{irr}$ is resonant to the magneto plasmon polariton dispersions. Hence, transport exhibits the same resonances at polariton frequencies as the THz-transmission experiment shown in Figs.~\ref{figure1}d). 
In contrast, at integer filling factors (bottom panel) we observe a set of linear dispersions, which are attributed to the excitation of an inter-Landau level transition and its higher orders. Their slopes are consistent with previous findings\cite{paravicini2017gate}. 
 
We understand the change of the longitudinal photo-response upon irradiation in the following way. First of all, we note that the absorption does not change with filling factor (as it is apparent in Fig.\ref{figure1}d). We also note that a simple thermal response will not explain our data, as the reduction of the maximum resistivity at the peak of the Shubnikov-de Haas oscillation is {\em not} accompanied by a concomitant increase at the minima. In a thermal response picture, the two color plots shown in Fig.~\ref{figure4}a) would give an inverted contrast of each other.

The experiment is compared to an extension of the theory \cite{bartolo2018vacuum} reported in the supplementary material. This theoretical approach accounts for a small polariton population inducing a change of the electronic scattering time.

As schematically shown in the upper (lower) panel of Fig.~\ref{figure4}b) the Fermi energy lies in the delocalized (localized) electronic states for half-integer (integer) filling factors.
Extended states - responsible for the polariton formation - have a spacial extention given by the magnetic length $l_0=\sqrt{c\hbar/(eB)}$ and hence a large dipole moment proportional to the square root of the filling factor ($d \sim e l_0 \sqrt{\nu}$).
In contrast, localized states exhibit a strongly reduced dipole matrix element due their localized nature. This is consistent with experimentally observed long excited state lifetimes \cite{kawano2001highly,arikawa2017light}. Exhibiting a greatly reduced dipole moment compared to extended states, localized states have a negligible overlap with the polariton wavefunction.

We therefore attribute the contrast seen in the lower panel of Fig.~\ref{figure4}a) to a second order process. This is also supported by the finding, that we observe higher order cyclotron excitations (black dashed lines), which cannot be directly excited by a photon due to optical selection rules. In agreement with the picture shown in the lower panel of Fig.~\ref{figure4}c), we observe that the photo-response maps the non-radiative decay channels of the polaritons into higher Landau levels. A polariton excitation (broad magenta line) can decay into the matter part and excite a localized electron from its ground state to a higher Landau Level (black dashed lines). This will occur most efficiently when the energy of the polariton matches that of the excited Landau level (marked with red ovals where this condition is met).

In summary, we measured the magneto-transport of a hybrid sample which integrates a Hall bar and a sub-wavelength metallic microwave resonator. Our results demonstrate a cavity quantum electrodynamic correction to magneto-transport. Mixed light-matter states - Landau-polaritons -  change the properties of the ground state of the electron gas. We observe a vacuum field induced change of magneto-transport. In a second set of experiments under weak single frequency microwave illumination, we reveal the signatures of the ultrastrong coupling regime by measuring the longitudinal resistance change of the two dimensional electron gas. Intriguingly, the response is strongly filling factor dependent up to filling factors of $\nu\sim50$. Transport through delocalized states in the center of a Landau Level are clearly subject to coupling to the cavities' vacuum field fluctuations. In contrast, localized states in the tails of each Landau Level appear to be protected from the severe modification of the cavity quantum electrodynamic environment. Transport in this regime maps the non-radiative decay channels of the polariton. The finding that magneto-transport carries signatures of the vacuum field fluctuations mediated by the polaritonic interaction paves the way towards vacuum-field-controlled many-body states in quantum Hall systems.

\subsection{\textit{Code availability}}
The code generated during the current study is available from the corresponding authors on reasonable request.

\subsection{\textit{Data availability}}
The datasets generated during and analysed during the current study are available from the corresponding authors on reasonable request.

\subsection{Acknowledgements}
We thank Peter M\"arki for the valuable contributions to the measurement electronics and Shima Rajabali for processing one of the transmission samples. The authors acknowledge financial support from the ERC grant No. 340975 (MUSiC). The authors also acknowledge financial support from the Swiss National Science Foundation (SNF) through the National Centre of Competence in Research Quantum Science and Technology (NCCR QSIT).

\section{Methods}
The sample is placed in a He3/He4-dilution fridge, in the center of a split-coil superconducting magnet in Helmholtz configuration with the sample surface perpendicular to the magnetic field axis. We obtain the difference frequency of two temperature tunable distributed diode feedback lasers from Toptica Photonics. This results in a single frequency source tunable from around 60 GHz to 600 GHz with an output power of 1 $\mu$W at 100 GHz and quickly decaying towards higher frequencies(see supplementary Fig.~S3). The divergent THz light source produced is collected with a parabolic mirror, passed through a HDPE window and focussed with a lens onto the sample. The resulting electron temperature is around 100 mK. A current of $I=$100 nA modulated at 14 Hz is applied from source to drain. The voltage difference $V_{xx}$ is measured using a commercial digital Lock-in from Zurich Instruments in conjunction with a homemade AC differential voltage pre-amplifier. For a 1 second integration time, we obtain a noise level of a few nV. A typical radiation induced resistance change of 1 $\Omega$ (see caption of Fig.~\ref{figure4}) corresponds to a signal of around 40 nV.

\end{document}